# Title: A Fermi-degenerate three-dimensional optical lattice clock

**Authors:**

S. L. Campbell[1,2]†, R. B. Hutson[1,2]†, G. E. Marti[1], A. Goban[1], N. Darkwah Oppong[1]‡, R. L. McNally[1,2]§, L. Sonderhouse[1,2], J. M. Robinson[1,2], W. Zhang[1]°, B. J. Bloom[1,2]•, J. Ye[1,2]*

**Affiliations:**

[1]JILA, NIST and University of Colorado, 440 UCB, Boulder, Colorado 80309, USA

[2]Department of Physics, University of Colorado, 390 UCB, Boulder, Colorado 80309, USA

†These Authors contributed equally to this work.

‡Present address: Max-Planck-Institut für Quantenoptik, Hans-Kopfermann-Straße 1, 85748 Garching, Germany

§Present address: Department of Physics, Columbia University, 538 West 120[th] Street, New York, New York 10027-5255, USA

•Present address: Rigetti Computing, 775 Heinz Ave, Berkeley, CA 94710, USA

°Present address: National Institute of Standards and Technology (NIST), 325 Broadway, Boulder, Colorado, 80305, USA

*To whom correspondence should be addressed; E-mail: ye@jila.colorado.edu

**Abstract**:

Strontium optical lattice clocks have the potential to simultaneously interrogate millions of atoms with a high spectroscopic quality factor of $4 \times 10^{17}$. Previously, atomic interactions have forced a compromise between clock stability, which benefits from a large atom number, and accuracy, which suffers from density-dependent frequency shifts. Here, we demonstrate a scalable solution which takes advantage of the high, correlated density of a degenerate Fermi gas in a three-dimensional optical lattice to guard against on-site interaction shifts. We show that contact interactions are resolved so that their contribution to clock shifts is orders of magnitude lower than in previous experiments. A synchronous clock comparison between two regions of the 3D lattice yields a $5 \times 10^{-19}$ measurement precision in 1 hour of averaging time.

**Main Text:**

Atomic clocks are advancing the frontier of measurement science, enabling tabletop searches for dark matter and physics beyond the Standard Model (*1–4*), as well as providing innovative quantum technologies for other branches of science (*5*).

One-dimensional (1D) optical lattice clocks provide a many-particle optical frequency reference that, together with advances in optical local oscillators, has led to record clock stability (*6, 7*) and the accompanying ability to evaluate systematic shifts to greater accuracy than ever before (*8–12*). However, the need to avoid atomic interactions places increasingly challenging engineering constraints on 1D optical lattice systems (*13–15*). A three-dimensional (3D) optical lattice was used in (*16*) to suppress atomic interactions in a clock based on a thermal gas of bosons. As the next step, the tools developed for quantum gas research (*17–20*) can be used to prepare metrologically useful quantum matter, combining the benefits of single-particle quantum state control and correlated many-particle systems.

Here, we use these tools to implement a 3D optical lattice clock with a degenerate Fermi gas. In this configuration, the atom number can in principle be scaled by orders of magnitude, while strong interactions prevent systematic errors associated with high atomic density. Specifically, we load a two-spin degenerate Fermi gas into the ground band of a 3D optical lattice in the Mott-insulating regime where interactions are responsible for a suppression of doubly occupied sites (*18, 29, 21*). This enables us to maximize atomic density while greatly suppressing collisional frequency shifts. For our coldest samples, the number of doubly occupied sites is suppressed by orders of magnitude as compared to the expected value for a non-interacting gas (*22*). For any residual atoms in doubly occupied sites, the enhanced interaction energy in a 3D lattice ensures that their transitions are well-resolved from the unperturbed clock transitions for atoms in singly occupied sites. Consequently, we are able to operate at a density above $10^{13}$ atoms/cm$^3$, three orders of magnitude greater than in previous lattice clock experiments (*10*).

Laser cooling followed by optical pumping to the $m_F = \pm 9/2$ stretched nuclear spin states produces a two-spin Fermi gas with an initial phase space density of 0.1 in a crossed optical dipole trap (XODT). Evaporative cooling to degeneracy proceeds by exponentially decreasing the trap depth in a 7 s, two-stage ramp (*23,24*). For different measurement goals, we optimize particular final parameters such as temperature (10 to 60 nK) and atom number ($10^4$ to $10^5$). The temperature $T$ and Fermi temperature $T_F$ are determined from a fit of a freely expanding gas to the Fermi-Dirac distribution, giving $T/T_F = 0.2 - 0.3$.

The atoms are then adiabatically loaded from the XODT into the ground band of a 3D optical lattice. We experimentally verify adiabaticity by measuring that $T/T_F$ does not increase by more than 10% after round trip loading back into the XODT. As the lattice depth rises, the increasing role of interactions relative to tunneling suppresses multiple occupancies in the Mott insulating regime. At the final lattice depths of 40 to 100 $E_{\text{rec}}$, where $E_{\text{rec}}$ is the lattice photon recoil energy, the Lamb-Dicke requirement is satisfied for clock light along all directions (*25*). Spectroscopy is performed on the 698 nm $^1S_0$ ($|g; m_F\rangle$) $\leftrightarrow$ $^3P_0$ ($|e; m_F\rangle$) clock transition. The clock laser propagating along the $\hat{x}$ lattice beam is used for precision spectroscopy (Fig. 1A), whereas an oblique clock laser enables a systematic characterization of the lattice via motional sideband spectroscopy (Fig. 1B). The absence of observable red-detuned sidebands demonstrates that the atoms are predominantly loaded into the ground band of the 3D lattice.

There has been a long-standing question as to whether the overall ac Stark shift in a 3D lattice can be managed to allow state-of-the-art narrow line clock spectroscopy. We implement a solution to this challenge, inspired by the proposal in (*26*). The differential ac Stark shift from the lattice trapping beams at a particular trap depth $\mathcal{U}_0$ can be expressed in terms of its scalar, vector, and tensor components as (*26, 27*),

$$\Delta\nu = (\Delta\kappa^s + \Delta\kappa^v m_F \xi \hat{e}_k \cdot \hat{e}_B + \Delta\kappa^t \beta)\mathcal{U}_0 \tag{1}$$

where $\Delta\kappa^{s,v,t}$ are the scalar, vector, and tensor shift coefficients, respectively, $\xi$ is the lattice light ellipticity, and $\hat{e}_k$ and $\hat{e}_B$ are unit vectors along the lattice beam wave vector and magnetic field quantization axis, respectively. The parameter $\beta$ can be expressed as $\beta = (3\cos^2\theta - 1)[3m_F^2 - F(F+1)]$, where $\theta$ is the angle between the nearly linear lattice polarization and $\hat{e}_B$.

We achieve state-independent trapping by operating the lattice at the combined scalar and tensor magic frequency and ensuring that the vector shift is zero (*10, 28*). Linearly polarized lattice light ($\xi = 0$) suppresses the vector shift, and the tensor shift is minimally sensitive to drifts in $\theta$ when the polarization is either parallel ($\theta = 0°$) or perpendicular ($\theta = 90°$) to the quantization axis. The frequency of the trapping light is then tuned to adjust the scalar shift so that it precisely cancels the tensor component. The $\theta = 0°$ configuration has been thoroughly studied in 1D lattice clocks (*10, 28*). For the 3D lattice, we set the horizontal ($\hat{x}, \hat{y}$) and vertical ($\hat{z}$) lattice polarizations to be parallel and perpendicular to $\hat{e}_B$, respectively (Fig. 1A). The two polarization configurations have distinct magic frequencies owing to their different tensor shifts.

We measure the magic frequencies for the vertical and horizontal lattice beams. A continuous wave Ti:Sapphire laser is used for the lattice light because of its low incoherent background (*29*). The absolute frequency of the lattice laser is traceable to the UTC NIST timescale through an optical frequency comb. For a given lattice laser frequency, we measure the differential ac Stark shift using four interleaved digital servos that lock the clock laser frequency to the atomic resonance for alternating high and low lattice intensities and $m_F = \pm 9/2$ spin states (*8, 10, 28*). From the data shown in Fig. 1C, we measure that the vertical ($\theta = 90°$) and horizontal ($\theta = 0°$) magic frequencies are 368.554839(5) THz and 368.554499(8) THz, respectively, in agreement with (*10, 28, 29*). From these two magic frequencies, we find that the scalar magic frequency is 368.554726(4) THz, in agreement with (*29*).

Because there are only two $\theta$ configurations with a stable tensor shift, in a 3D lattice two of the three lattice beams will necessarily have the same magic frequency. However, the frequencies of all lattice beams must be offset to avoid interference, which is known to cause heating in ultracold quantum gas experiments (*30*). We choose our two horizontal beams to have the same polarization and operate them with equal and opposite detunings ($\pm 2.5$ MHz) from their magic frequency, giving equal and opposite ac Stark shifts from the two beams. From the slopes in Fig. 1C, we determine that for a 10% imbalance in trap depths, detuning the two horizontal beams $\pm 2.5$ MHz from their magic frequency results in a $< 1 \times 10^{-18}$ systematic shift, the exact magnitude of which can be measured to much better accuracy.

Another issue resulting from our 3D geometry is that one lattice beam must operate with $\hat{e}_k \cdot \hat{e}_B = 1$, which can give rise to a vector ac Stark shift owing to residual circular polarization. The vertical ($\hat{z}$) beam has this configuration, and we measure a $3 \times 10^{-18} \times m_F/E_{\text{rec}}$ vector shift (*22*), corresponding to an ellipticity of $\xi = 0.007$ (*26*). In contrast, the horizontal beams ($\hat{x}, \hat{y}$) with $\hat{e}_k \cdot \hat{e}_B = 0$ enjoy an additional level of vector shift suppression. Because the clock operates by locking to alternating opposite spin states, the net vector shift can be removed.

In the ground band of the lattice, each site can be occupied by either one atom, or by two atoms with opposite nuclear spin. Tight confinement in the 3D lattice gives rise to strong interactions on doubly occupied sites. We label the two-particle eigenstates of the two-orbital interaction Hamiltonian as $|gg; m_F, m_F'\rangle = |gg\rangle \otimes |s\rangle$, $|eg^+; m_F, m_F'\rangle = (|eg\rangle + |ge\rangle)/\sqrt{2} \otimes |s\rangle$, $|eg^-; m_F, m_F'\rangle = (|eg\rangle - |ge\rangle)/\sqrt{2} \otimes |t\rangle$, and $|ee; m_F, m_F'\rangle = |ee\rangle \otimes |s\rangle$ with corresponding energies $U_{gg}$, $U_{eg^+}$, $U_{eg^-}$, and $U_{ee}$ (*31–34*). Here, $|s\rangle$ and $|t\rangle$ represent the singlet and triplet wavefunctions for the two spins, $m_F$ and $m_F'$. An applied magnetic bias field mixes the $|eg^+; m_F, m_F'\rangle$ and $|eg^-; m_F, m_F'\rangle$ states owing to a difference between the Landé *g*-factors of the two orbitals. We label the new eigenstates of the combined interaction and Zeeman Hamiltonian $|eg^u; m_F, m_F'\rangle$ and $|eg^d; m_F, m_F'\rangle$.

Clock light resonantly couples the ground state $|gg; m_F, m_F'\rangle$ only to the states $|eg^{u,d}; m_F, m_F'\rangle$. The energies $U_{eg^+}$ and $U_{eg^-}$ differ from $U_{gg}$ on the $h \cdot$ kHz scale, where $h$ is the Planck constant, resulting in transitions on doubly occupied sites that are well-resolved from the single-atom clock transitions and strong suppression of two-photon transitions to the $|ee; m_F, m_F'\rangle$ state. In 1D optical lattice clocks, interaction shifts are less than clock transition Rabi frequencies; in 2D clocks, interaction shifts and Rabi frequencies are comparable (*35*). In contrast, the interaction shifts in a 3D lattice are $10^3$ times greater than the clock transition Rabi frequency. Although for normal clock operation we load at most one atom per site, to measure contact interactions we increase the final Fermi temperature so that a small fraction of lattice sites are filled with two atoms in the $|gg; m_F, m_F'\rangle$ state. Figure 2A shows clock spectroscopy for a magnetic field $B = 500$ mG. At this field, $\pi$-polarized clock light gives a negligible transition amplitude to $|eg^u; m_F, m_F'\rangle$ state thanks to destructive interference between the two oppositely signed transition dipole moments for the stretched states. Therefore, we observe excitation to the $|eg^d; m_F, m_F'\rangle$ state, highlighted in red, and no excitation to the $|eg^u; m_F, m_F'\rangle$ state, highlighted in blue.

Clock operation predominantly probes atoms in singly occupied sites, with only a negligible systematic shift due to line pulling caused by transitions from the few atoms in doubly occupied sites. For a comprehensive study of line pulling from doubly occupied sites, we use our measurements of $U_{eg^-} - U_{gg}$ and $U_{eg^+} - U_{gg}$ at $B = 0$ to calculate the spectrum of all transitions on doubly occupied sites as a function of $B$ (Fig. 2B) (22). We account for imperfect clock laser polarization and residual spin populations by considering $\sigma^-$, $\sigma^+$ and $\pi$ transitions to all $|eg^{u,d}; m_F, m_F'\rangle$ states, as indicated in the shaded regions in Fig. 2B. At intermediate magnetic fields, transitions from doubly occupied sites can cross clock transitions and potentially cause significant frequency shifts. However, this is avoided if $B$ is kept sufficiently low. At our chosen bias field of 500 mG (8, 10), transitions on doubly occupied sites are at least 500 Hz away. From a 1% upper bound on residual transition amplitudes, we estimate the fractional frequency shift caused by line pulling from doubly occupied sites to be below $1 \times 10^{-24}$ for 1 Hz linewidths (22). We also calculate the fractional frequency shifts coming from superexchange interactions (36) between neighboring sites at typical lattice depths and find them to be below $1 \times 10^{-22}$.

Although on-site interaction shifts can be eliminated in 3D lattice clocks, atoms may also interact via long-range electric dipole forces, which can lead to many-body effects such as collective frequency shifts, superradiance, and subradiance (37). At unit filling, clock shifts from dipolar interactions could reach the $10^{-18}$ level though employing different 3D lattice geometries is a promising approach for canceling these shifts (38, 39). Exploring different strategies for accurately measuring and eliminating dipolar frequency shifts will be a fruitful avenue of future study.

With atomic interactions and lattice ac Stark shifts controlled, we demonstrate the longest coherence times in atom-light interactions. Figure 3 shows a progression of Ramsey fringes with free evolution times from 100 ms to 6 s, beyond what has been demonstrated in 1D lattice clocks (*10, 15, 40*). The $\hat{x}$ lattice beam is operated at a depth sufficient ($> 80\ E_{\text{rec}}$) to prevent atoms from tunneling along the clock laser axis during the 6 s free evolution period. Spectroscopy is performed on a spin-polarized sample, prepared by first exciting $|g; -9/2\rangle \rightarrow |e; -9/2\rangle$ then removing all ground state atoms via resonant $^1S_0 - {}^1P_1$ light. Our longest observed coherence time approaches the limit of our clock laser based on its noise model (*41*) and the 12.7 s dead time between measurements. For the demonstration of Ramsey fringes at longer free-evolution times, maintaining atom-light phase coherence will require a significant reduction in fundamental thermal noise from the optical local oscillator. Additionally, the observation of narrower lines will require magnetic field control below the 100 μG level. The contrast of the observed Ramsey fringes is likely limited by lattice light causing both dephasing over the atomic sample and excited state population decay. The external harmonic confinement of the lattice beams limits the size of the Mott insulating region in the center of the trap to ~$2 \times 10^4$ atoms. We maximize the number of singly occupied sites in the center of trap by operating with $1 \times 10^4$ atoms at a temperature of 15 nK.

The combination of large atom numbers with a long atom-light coherence time enabled by this system opens the possibility for improving the quantum projection noise (QPN) limit to clock stability by more than an order of magnitude over the current state-of-the-art (*10,40*). The QPN limit for Ramsey spectroscopy is,

$$\sigma_{\text{QPN}}(\tau) = \frac{1}{2\pi\nu T}\sqrt{\frac{T+T_d}{N\tau}}, \tag{2}$$

where $\nu$ is the clock frequency, $T$ is the free-evolution time, $T_\text{d}$ is the dead time, and $\tau$ is the total averaging time. Typically, optical clocks operate at a stability above this limit owing to the Dick effect (*42*). However, operation at or near the QPN limit has recently been demonstrated in a number of different systems; common mode rejection of clock laser noise has been achieved through synchronous interrogation of two independent atomic ensembles (*6, 43*) and two separated ions in a single ion trap (*44*), whereas interleaved interrogation of two clocks with zero dead time has enabled continuous monitoring of the clock laser phase (*40*).

Here, we demonstrate synchronous clock comparison between two spatially separated atomic ensembles in the 3D lattice. The regions, represented by the blue and orange spheres in Fig. 1A, are rectangular cross sections separated from one another by > 6 µm, each containing 3000 atoms. After a $T = 4$ s Ramsey sequence, we perform absorption imaging on the $^1S_0 - {^1P_1}$ transition with a 1 µm imaging resolution to independently measure the excitation fractions $P_1$ and $P_2$ in the two respective regions. Although most clock shifts are common-mode suppressed between the two regions, the slightly elliptical polarization of the $\hat{z}$ lattice beam produces a few mHz differential vector Stark shift between the two regions. We measure this effect by probing the $|g; 1/2\rangle \rightarrow |e; 1/2\rangle$ clock transition. The shift manifests itself as a fixed phase between the signals $P_1$ and $P_2$. Plotting $P_1$ vs. $P_2$ (Fig. 4A, inset) allows us to extract this differential shift in a manner independent of the clock laser noise through either ellipse fitting (*45*) or Bayesian estimation (*46*). The QPN is uncorrelated between the two independently detected regions and thus limits the precision of the extracted phase shift. In 2.2 hours of averaging, we measure a 5.56(15) mHz shift, corresponding to an instability of $3.1 \times 10^{-17}/\sqrt{\tau}$ (Fig. 4A) and a measurement precision of $3.5 \times 10^{-19}$ (Fig. 4B). To observe the scaling of measurement precision with atom number, we reduce the atom number by a factor of three and observe an increase in QPN noise by $\sqrt{3}$ (red dashed line in Fig. 4A).

By using larger lattice beam waists, our design can accommodate even greater atom numbers, which, combined with reduced preparation time of degenerate gases and increased atomic coherence in the lattice, should enable the operation of synchronous comparisons better than the $10^{-18}/\sqrt{\tau}$ level, leading to a new generation of precision measurement tools, including space-based gravitational wave detectors (*47*). Reaching such performance is extremely challenging for 1D OLCs as collisional effects force a compromise between interrogation time and the number of atoms that can be simultaneously interrogated (*10, 15*).

With quantum degenerate atoms frozen into a 3D lattice, we can further advance the state-of-the-art in coherent atom-light interactions with the next generations of ultrastable optical reference cavities based on crystalline materials (*48–50*). Quantum degenerate clocks also provide a promising platform for studying many-body physics. Future studies of dipolar interactions will not only be necessary for clock accuracy, but will also provide insight into long-range quantum spin systems in a regime distinct from those explored by polar molecules (*51, 52*), Rydberg gases (*53, 54*), and highly magnetic atoms (*55–58*). When clocks ultimately confront the natural linewidth of the atomic frequency reference, degenerate Fermi gases may be useful for engineering longer coherence times through Pauli blocking of spontaneous emission (*59*) or collective radiative effects (*39, 60*).

**Acknowledgments:**

We thank D.E. Chang and H. Ritsch for insightful discussions on dipolar interactions in a 3D lattice. We also acknowledge technical contributions and discussions from C. Benko, T. Bothwell, S. L. Bromley, K. Hagen, J. L. Hall, B. Horner, H. Johnson, T. Keep, S. Kolkowitz, J. Levine, T. H. Loftus, T. L. Nicholson, E. Oelker, D. G. Reed, and X. Zhang. This work is supported by NIST, DARPA, AFOSR-MURI, and the NSF JILA Physics Frontier Center. G.E.M. is supported by a National Research Council Postdoctoral Fellowship, A.G. is supported by a JSPS Fellowship, and L.S. is supported by a NDSEG fellowship.

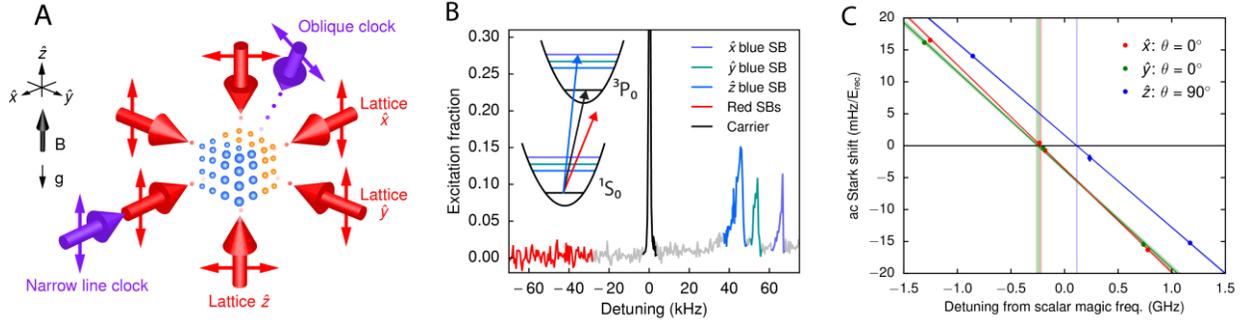

**Fig. 1**. **A Fermi-degenerate 3D optical lattice clock.** (**A**) Schematic showing propagation direction (large arrows) and polarization (double arrows) of the 3D lattice and clock laser beams. The quantization axis is defined by the magnetic field $B$. The narrow line clock laser used for precision spectroscopy is phase-stabilized to lattice $\hat{x}$. The oblique clock laser is used to drive motional sidebands along all three lattice axes. (**B**) Motional sideband spectroscopy using the oblique clock laser shows no observable red sidebands, illustrating that atoms are predominantly in the ground band of the lattice. (**C**) Determination of the magic wavelengths for the horizontal ($\hat{x}$, $\hat{y}$) and vertical ($\hat{z}$) lattices, with $m_F = \pm 9/2$. The measured frequency shift is scaled by the difference in trap depths between the high and low lattice intensities (*22*). The difference in slopes is caused by trapping potential inhomogeneities that do not affect the determination of the magic frequencies.

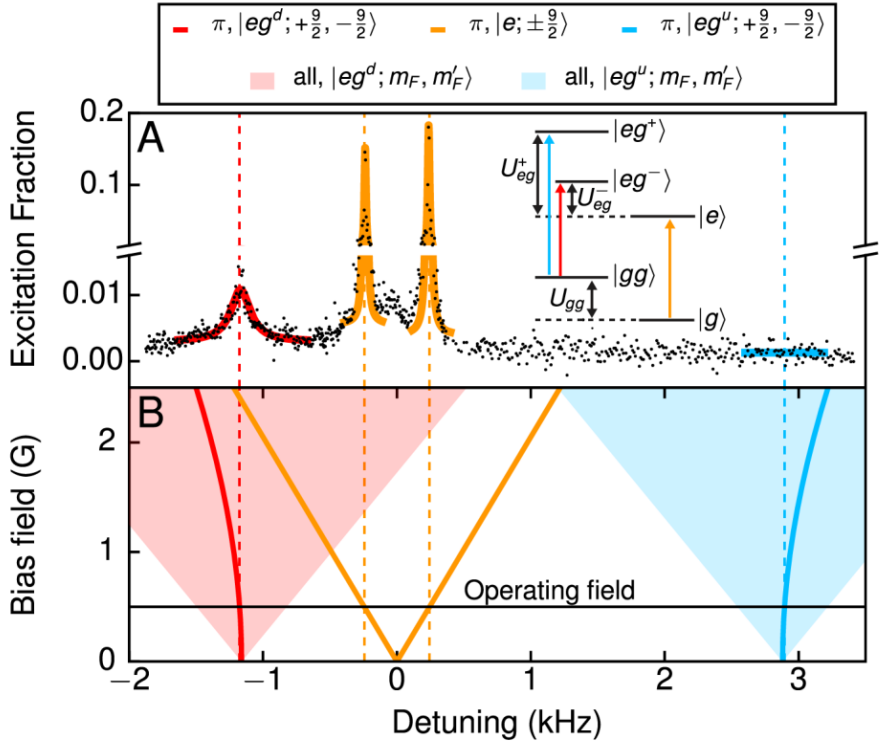

**Fig. 2. Resolved atomic contact interactions**. (**A**) Clock spectroscopy data of a two-spin Fermi gas in the $m_F = \pm 9/2$ stretched states for a 500 mG magnetic field, where a small fraction of the lattice sites contain both spin states. All transitions are saturated. The $|gg; m_F, m'_F\rangle \to |eg^u; m_F, m'_F\rangle$ transition is absent owing to its vanishing dipole matrix element at small magnetic fields. Inset: Level diagram at zero magnetic field. (**B**) Calculated detunings for transitions on singly and doubly occupied sites (*22*). The solid lines correspond to transitions on singly occupied (orange) and doubly occupied (red, blue) sites with $m_F = \pm 9/2$. Transitions on doubly occupied sites for arbitrary $m_F$ and clock laser polarization lie within the shaded regions. At our operating magnetic field of 500 mG, all resonances for doubly occupied sites are well-resolved from the clock transitions.

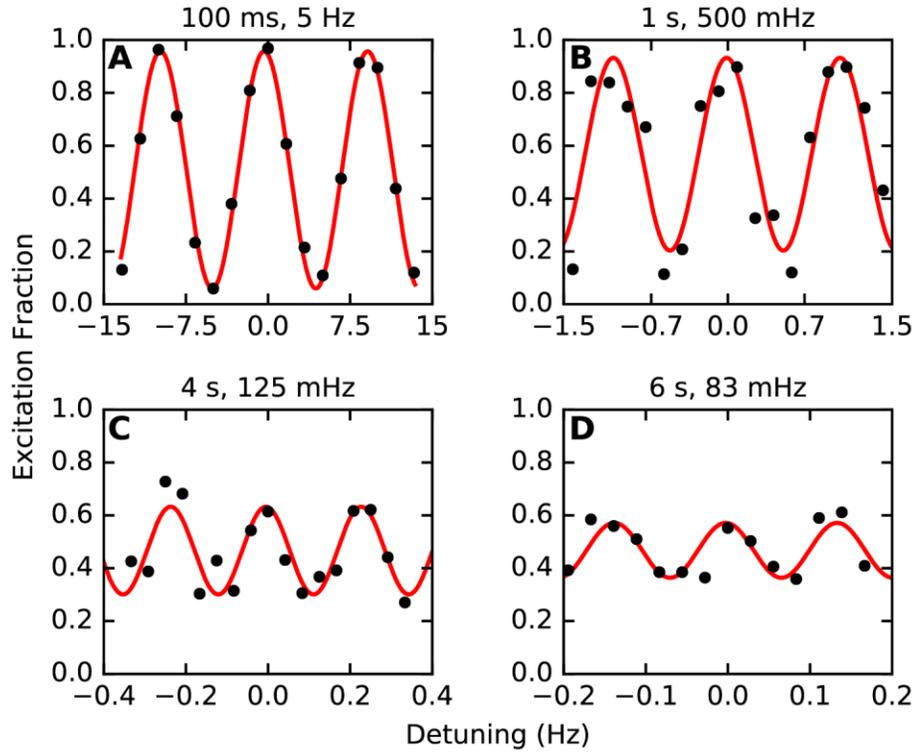

**Fig. 3**. **Narrow-line clock spectroscopy**. Ramsey spectroscopy data taken with $1 \times 10^4$ atoms at 15 nK for (**A**) 100 ms, (**B**) 1 s, (**C**) 4 s, and (**D**) 6 s free-evolution times, using 10 ms π/2 pulse times. With contact interactions and ac Stark shifts controlled in a 3D lattice, we are able to measure fringes at an unprecedented 6 s free-evolution time with a density of over $10^{13}$ atoms/cm$^3$.

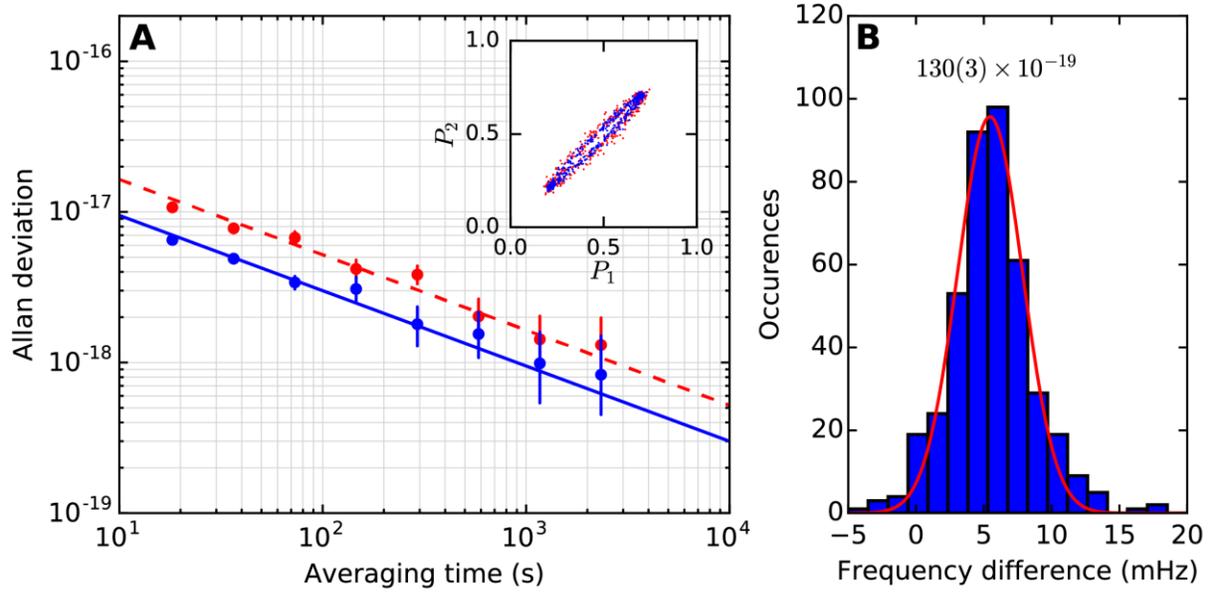

**Fig. 4**. **Synchronous clock comparison.** (**A**) The Allan deviation of the differential frequency shift between two independent regions of the 3D lattice demonstrates a $3.1 \times 10^{-17}/\sqrt{\tau}$ instability for 3000 atoms in each region (data in blue filled circles, fit in blue solid line). The frequency difference is determined through a Bayesian estimation algorithm that determines the eccentricity of the ellipse of the parametric plot of $P_1$ vs. $P_2$ (inset, 420 runs). When the atom number is reduced to 1000 in each region, the instability increases by $\sqrt{3}$ (red filled circles, red dashed line). The error bars represent 95% confidence intervals, assuming white frequency noise. (**B**) A histogram of the measured frequency differences for an averaging time of 2.2 hours and atom number of 3000. The fitted Gaussian gives a fractional frequency difference of $129.6(3.5) \times 10^{-19}$.

SUPPLEMENTARY MATERIALS
Material and Methods
Tables S1 and S2
Figs. S1 to S6
References (*61-63*)

# Supplementary Materials

## Materials and Methods

In an ultrahigh vacuum chamber with a >90 s background gas collision-limited lifetime, we perform initial laser cooling of $^{87}$Sr in a magneto-optical trap (MOT) on the 30 MHz $^1S_0 - {}^1P_1$ transition. We then load atoms into a second-stage MOT which operates on the 7 kHz $^1S_0 - {}^3P_1$ intercombination transition. For evaporative cooling, the atoms are next loaded into a 1064 nm crossed optical dipole trap (XODT), which is formed by a horizontal and a vertical beam crossing at their respective waists. The XODT is held on throughout both the broadband and narrow-line cooling stages in the second-stage MOT. To ensure that narrow-line cooling works efficiently inside the XODT, the spatially dependent differential ac Stark shifts of the $^1S_0 - {}^3P_1$ cooling transition must be minimized.

The horizontal beam has a sheetlike geometry with waists of 340 $\mu$m and 17 $\mu$m in the horizontal and vertical directions, respectively. By focusing more tightly in the vertical direction, the horizontal beam achieves the same confinement against gravity for a lower light intensity. The vertical beam, which is aligned at a small angle with respect to gravity, is a circular Gaussian beam with a 64 $\mu$m beam waist. Similar to (*24*), the vertical beam is circularly polarized to minimize differential ac Stark shifts. We perform a further stage of cooling by blue-detuning the cooling light relative to the free space resonance to address atoms inside the XODT.

After the MOT is switched off, $5 \times 10^6$ atoms at 1.5 $\mu$K are loaded into the XODT, in an equal mixture of all 10 nuclear spin states. To prepare an equal mixture of the stretched $m_F = \pm 9/2$ states, we perform optical pumping via the $^1S_0 \leftrightarrow {}^3P_1, F = 9/2$ transition in a 3 G magnetic bias field, which splits neighboring $m_F$ states by 260 kHz. First, a $\sigma^-$-polarized optical pumping beam is frequency-chirped from the $m_F = 1/2 \to 3/2$ transition to the $m_F = 7/2 \to 9/2$ transition, which pumps all $m_F < 0$ atoms into the $m_F = -9/2$ state.

In the second step, a liquid crystal waveplate is used to switch the laser polarization to $\sigma^+$, and similarly all the $m_F > 0$ atoms are pumped into the $m_F = 9/2$ state.

## 3D Lattice Design

First, we consider both the requirements for loading into the ground band of the lattice and for reaching the Mott-insulating regime. Loading the ground band requires that $E_{F,\text{ODT}}, k_B T_\text{lattice} \ll E_\text{rec}$, where $E_{F,\text{ODT}}$ is the Fermi energy in the ODT, $k_B$ is the Boltzmann constant, $T_\text{ODT}$ is the temperature in the ODT, and $E_\text{rec}$ is the recoil energy from a lattice photon. Competition between tunneling ($J$) and repulsive interactions ($U$) initializes the spatial distribution of the atoms. As the lattice depth increases, multiple occupancies are suppressed when $12J \ll U$ and $k_B T_\text{lattice} \ll E_F \ll U$, where $T_\text{lattice}$ is the temperature in the lattice (*18, 19*).

We verify lattice loading adiabaticity by measuring the $T/T_F$ both before ramping up the lattice and after a subsequent reversed ramp back into the XODT. For the conditions of $N = 10^5$ and $T = 50$ nK, we observe no increase for $T/T_F = 0.30 \pm 0.05$, which allows us to put a conservative upper bound on the bulk gas entropy for the data in Figs. 1 and 2. The data in Fig. 3 and 4 were taken under different conditions: $N = 10^4$, $T = 15$ nK and $T/T_F = 0.2$. We have verified, again through a round-trip measurement of $T/T_F$, that loading of such a sample to the lattice is nearly adiabatic. We detected less than a few nK increase in temperature and a 30% loss in atom number, corresponding to a increase in $T/T_F$ of 10%.

To estimate suppression of doubly-occupied sites (doublons) in the lattice due to interactions, we use a model which assumes conservation of the total entropy and atom number measured in the XODT to compare the doublon fraction for the interacting and non-interacting cases (*18, 19*). For our coldest samples, a large suppression of doublons, as well as a central region of vanishing compressibility, is expected (see Fig. S1). The model predicts a dublon fraction more than 100 times smaller than that of the non-interacting case. Clock spectroscopy comfirms the suppression of doublons. The next step is to use a high-resolution imaging objective to verify the vanishing compressibility in the center of the trap and the existence of a low-entropy Mott-insulator (*18, 19*).

Next, we consider how finite tunneling rates affect clock spectroscopy. We require $1/J_x \gg \tau$, where $J_x$ is the tunneling rate along the clock laser propagation direction and $\tau$ is the spectroscopy time, as the finite Bloch bandwidth of the lattice potential causes a first-order Doppler broadening of $8J_x$. This requirement is satisfied for our longest spectroscopy times $\tau$ = 6 s by using lattice depths above $80E_{\text{rec}}$. To achieve a sufficiently deep trap as well as mode-match with the XODT, we use elliptical beams for the $\hat{x}$ and $\hat{y}$ lattice axes with horizontal and vertical waists of 120 $\mu$m and 35 $\mu$m, respectively. The $\hat{z}$ lattice beam is round with a 90 $\mu$m waist. Typical operating parameters in our experiment are summarized in Table S1, and a list of the requirements on these experimental parameters is summarized in Table S2.

## Interactions and Line Pulling

We calculate the detunings of transitions from doubly occupied sites, relative to the single-atom clock transition $|g; m_F\rangle \rightarrow |e; m_F\rangle$ at zero magnetic field. Since the clock operates under a magnetic bias field $B$, we consider how competing Zeeman and interaction energies determine the energy eigenstates for atoms on doubly occupied sites with arbitrary $m_1$ and $m_2$. The Zeeman shifts for $|g; m_F\rangle$ and $|e; m_F\rangle$ are $g_I \mu_B B m_F$ and $(g_I + \delta g)\mu_B B m_F$, where $g_I$ is the nuclear $g$-factor ($g_I \mu_B = h \cdot 184.4$ Hz/Gauss) and $\delta g$ is the differential $g$-factor between the ground and clock states ($\delta g \mu_B = h \cdot 108.4$ Hz/Gauss). The differential Zeeman shift between the two clock states introduces a coupling between the $|eg^+; m_1, m_2\rangle$ and $|eg^-; m_1, m_2\rangle$ states. Thus, the combined interaction and differential Zeeman Hamiltonian in the $|eg^\pm; m_1, m_2\rangle$ basis can be expressed as (*32, 33*):

$$\hat{H} = \begin{pmatrix} U_{eg}^+ + \frac{\Delta_1+\Delta_2}{2} & \frac{\Delta_1-\Delta_2}{2} \\ \frac{\Delta_1-\Delta_2}{2} & U_{eg}^- + \frac{\Delta_1+\Delta_2}{2} \end{pmatrix}, \tag{S1}$$

where $\Delta_{1,2} = \delta g \mu_B B m_{1,2}$ are the differential Zeeman shifts. The two eigenstates $|eg^u; m_1, m_2\rangle$ and $|eg^d; m_1, m_2\rangle$ have eigenenergies given by,

$$E_{\{u,d\}}(m_1, m_2) = V + \frac{\Delta_1+\Delta_2}{2} \pm \sqrt{V_{\text{ex}}^2 + \left(\frac{\Delta_1-\Delta_2}{2}\right)^2}, \tag{S2}$$

where $V = \left(U_{eg}^+ + U_{eg}^-\right)/2$ and $V_{\text{ex}} = \left(U_{eg}^+ - U_{eg}^-\right)/2$ are the direct and exchange interaction energies, respectively. The detunings of the $|gg; m_1, m_2\rangle \to |eg^{\{u,d\}}; m_1', m_2'\rangle$ transitions relative to the $B=0$ single-atom clock transition are,

$$\Delta E_{\{u,d\}}(m_1, m_2; m_1', m_2') = E_{\{u,d\}}(m_1', m_2') - U_{gg} + g_I \mu_B B (m_1' + m_2' - m_1 - m_2), \tag{S3}$$

when the two-particle Rabi couplings $\Omega^{\{u,d\}}(m_1, m_2; m_1', m_2'; q)$ with $\pi(q=0)$ or $\sigma^\pm(q=\mp 1)$-polarized clock light are non-zero, as discussed below.

The atom-light Hamiltonian for a single particle driven by $\pi(\sigma^\pm)$-polarized clock light is given by,

$$\hat{H}_q^{(1)} = \frac{\hbar}{2} \sum_{m_F} \left(\Omega_{m_F, q} |e; m_F - q\rangle\langle g; m_F| + \text{h.c.}\right), \tag{S4}$$

Here, the single-particle coupling $\Omega_{m_F, q}$ is expressed using the Wigner-Eckart theorem as,

$$\Omega_{m_F, q} = -\langle F_g, m_F; 1, q | F_e, m_F - q\rangle \langle F_g ||d|| F_e\rangle \cdot \frac{\mathcal{E}_q}{\hbar}, \tag{S5}$$

where $\langle F_g ||d|| F_e\rangle$ is the reduced matrix element and $\mathcal{E}_q$ is the electric field amplitude. Then, the two-particle atom-light Hamiltonian is written as,

$$\hat{H}_q^{(2)} = \left(\hat{H}_q^{(1)}\right)_1 \otimes (\mathbb{1})_2 + (\mathbb{1})_1 \otimes \left(\hat{H}_q^{(1)}\right)_2. \tag{S6}$$

We find the two-particle couplings $\Omega^\pm(m_1, m_2; m_1', m_2'; q)$ between the $|gg; m_1, m_2\rangle$ and $|eg^\pm; m_1', m_2'\rangle$ transitions, driven by $\pi(q=0)$- or $\sigma^\pm(q=\mp 1)$-polarized clock beams to be,

$$\begin{aligned}
\frac{\hbar}{2}\Omega^\pm(m_1, m_2; m_1', m_2'; q) &= \langle eg^\pm; m_1', m_2'|\hat{H}_q^{(2)}|gg; m_1, m_2\rangle \\
&= \frac{\hbar}{2\sqrt{2}}\left[\Omega_{m_1,q}\left(\delta_{m_1',m_1-q}\delta_{m_2',m_2} \mp \delta_{m_1',m_2}\delta_{m_2',m_1-q}\right)\right.\\
&\quad \left. +\Omega_{m_2,q}\left(\pm\delta_{m_1',m_1}\delta_{m_2',m_2-q} - \delta_{m_1',m_2-q}\delta_{m_2',m_1}\right)\right]. \quad (S7)
\end{aligned}$$

Thus, the two-particle couplings $\Omega^{\{u,d\}}(m_1, m_2; m_1', m_2'; q)$ between the $|gg; m_1, m_2\rangle$ and $|eg^{\{u,d\}}; m_1', m_2'\rangle$ states at a given $B$ are obtained from linear combinations of Eq. S7. Fig. S4 shows all the detunings $\Delta E_{\{u,d\}}(m_1, m_2; m_1', m_2')$ from Eq. S3 for non-zero $\Omega^{\{u,d\}}(m_1, m_2; m_1', m_2'; q)$ with $\pi$ (black lines), $\sigma^+$ (red lines), and $\sigma^-$ (blue lines) clock laser polarizations, as well as all single-atom transitions. In the case of an equal mixture of $m_F = \pm 9/2$ stretched states, the $\pi$-polarized clock light does not drive the $|gg; 9/2, -9/2\rangle \to |eg^+; 9/2, -9/2\rangle$ transition, since $\Omega_{m_1=9/2,q=0} = -\Omega_{m_2=-9/2,q=0}$ gives zero coupling strength as seen in Eq. S7. At our operating $B$, $|eg^u\rangle \approx |eg^+\rangle$, so $\pi$-polarized clock light only drives the $|gg; 9/2, -9/2\rangle \to |eg^d; 9/2, -9/2\rangle$ transition, as shown in Fig. 2A.

Next, we investigate line pulling effects due to the additional lines shown in Fig. S4. Single-particle line pulling effects have already been discussed in (*28*); here we focus on line pulling from transitions on doubly occupied sites. We approximate the lineshapes as Lorentzians with full width at half maximum $\Gamma$. For typical clock operation, the error signal $\varepsilon$ for the digital PID that steers the clock laser center frequency $f_0$ to the atomic resonance is generated by measuring the normalized excitation fraction $N_{\text{exc}}$ at $f_0 + \Gamma/2$ and at $f_0 - \Gamma/2$ (*8, 10*). The clock frequency is locked to the atomic reference such that $\varepsilon = N_{\text{exc}}(f_0 - \Gamma/2) - N_{\text{exc}}(f_0 + \Gamma/2) = 0$. In Fig. S5, we study how an additional line with a given amplitude relative to the clock transition modifies the center frequency $f_0$ for which $\varepsilon = 0$. Transitions from atoms on doubly occupied sites begin to overlap with the clock transitions when $B > (U_{gg} - U_{eg}^-)/[(9\delta g + g_I)\mu_B]$. At our chosen bias field of 500 mG (*8, 10*), the transitions on doubly occupied sites are separated from the main clock transitions by at least 500 Hz. Using a 1% upper bound on residual transition amplitudes and assuming 1 Hz transition linewidths, we estimate the fractional frequency shift due to line pulling effects from doubly occupied sites to be below $1 \times 10^{-24}$.

## 3D Magic Wavelength

We evaluate the ac Stark shift using four interleaved but independent servos locked to the $m_F = \pm 9/2$ transitions for each of the high and low lattice intensity configurations, with high and low lattice depths differing by 36(2) $E_R$, 35(2) $E_R$, and 37(2) $E_R$ for the $\hat{x}$, $\hat{y}$, and $\hat{z}$ lattices, respectively. Each servo tracks the detuning between one transition of the strontium atoms and the TEM00 mode of an ultra-stable cavity (*15*).

Laser frequency drift gives a nonuniform shift to the interleaved locks that can lead to an erroneous offset in the measured ac Stark shift. Every four experimental cycles, we measure independent frequencies locked to the four configurations $\{f^i_{\text{high},+9/2}, f^i_{\text{low},+9/2}, f^i_{\text{high},-9/2}, f^i_{\text{low},-9/2}\}$, where high and low refer to lattice depths, $\pm 9/2$ refer to the $m_F$ states, and $i$ is the iteration number of the experiment. A four-point string analysis (*61, 62*) removes linear and quadratic laser drift using the following linear combination of eight consecutive measurements:

$$\Delta f^i_{\text{scalar+tensor}} = \frac{3}{16} \left( f^i_{\text{high},+9/2} + f^i_{\text{high},-9/2} \right) - \frac{5}{16} \left( f^i_{\text{low},+9/2} + f^i_{\text{low},-9/2} \right)$$
$$+ \frac{5}{16} \left( f^{i+1}_{\text{high},+9/2} + f^{i+1}_{\text{high},-9/2} \right) - \frac{3}{16} \left( f^{i+1}_{\text{low},+9/2} + f^{i+1}_{\text{low},-9/2} \right), \quad \text{(S8)}$$

$$\Delta f^i_{\text{vector}} = \frac{3}{16} \left( f^i_{\text{high},+9/2} - f^i_{\text{high},-9/2} \right) - \frac{5}{16} \left( f^i_{\text{low},+9/2} - f^i_{\text{low},-9/2} \right)$$
$$+ \frac{5}{16} \left( f^{i+1}_{\text{high},+9/2} - f^{i+1}_{\text{high},-9/2} \right) - \frac{3}{16} \left( f^{i+1}_{\text{low},+9/2} - f^{i+1}_{\text{low},-9/2} \right). \quad \text{(S9)}$$

The measured frequencies are correlated in time because the servos low-pass filter the system response. We correct the servo frequency using the error signal, which contains higher-frequency components of the system response, according to,

$$f_{\text{corrected}} = f_{\text{uncorrected}} + \frac{\Gamma}{2A} \times e, \tag{S10}$$

where $\Gamma$ is the FWHM linewidth, $e$ is the error signal (the difference between the excitation fraction from probing the left and right sides of the line), and $A$ is the maximum peak height of the spectroscopic feature. Fig. S2 demonstrates that this procedure flattens the autocorrelation function of the corrected data (blue) compared to the uncorrected data (green). To estimate the autocorrelation function from the data, we use the formula (*63*),

$$\rho_j = \frac{\sum_{i=1}^{n-j}[X_i - \bar{X}(n)][X_{i+j} - \bar{X}(n)]}{(n-j)S^2(n)}, \qquad S^2(n) = \frac{\sum_{i=1}^{n}[X_i - \bar{X}(n)]^2}{n(n-1)}. \tag{S11}$$

We use the autocorrelation function to calculate the unbiased error for correlated data,

$$\frac{\sigma}{\sqrt{n}}\sqrt{1 + 2\sum_{j=1}^{n-1}\left(1 - \frac{j}{n}\right)\rho_j}, \tag{S12}$$

where $\sigma$ is the standard deviation of the point-string difference frequencies $\Delta f^i$, $\rho_j$ is the autocorrelation function for data separated by $j$ points, and $n$ is the number of measured frequencies (*63*).

All of these steps are validated on simulated data modeled by clock laser noise similar to the actual laser (*41*), with a power spectral density $S \propto 1/f$, where $f$ is the frequency, and a linear frequency drift from imperfectly canceled material creep of the reference cavity between -2 and 2 mHz/s. We use weighted least-squares fitting to determine the magic wavelength for each lattice direction. From the difference in the splittings between the $\pm 9/2$ states for the high and low lattice intensities, we also extract the vector ac Stark shift, as shown in Fig. S3.

**Narrow line spectroscopy**

Using Rabi spectroscopy with a 4 s pulse time, we measure a full-contrast Fourier-limited linewidth of 190(20) mHz, as shown in Fig. S6. This is different than spectroscopy in a 1D lattice, where clock scans at the longest probe times show substantially reduced contrast due to atomic contact interactions (*10, 15*).

| Parameter | Symbol | Typical Value |
|---|---|---|
| Fermi energy in XODT | $E_{\text{F,XODT}}$ | 75 nK·$k_B$ |
| Temperature in XODT | $T_{\text{XODT}}$ | 15 nK |
| Lattice recoil energy | $E_{\text{rec}}$ | 167 nK·$k_B$ |
| Clock laser recoil energy | $E_{\text{rec,clock}}$ | 226 nK·$k_B$ |
| Contact interaction energy | $U$ | 10 kHz·$h$ |
| Tunneling rate along the clock laser axis | $J_x/h$ | 0.5 mHz |
| Superexchange interactions | $J^2/U$ | $10^{-8}$ Hz |
| Spectroscopy time | $\tau$ | 6 s |
| Lattice trap depth | $\mathcal{U}_{0,x}, \mathcal{U}_{0,y}, \mathcal{U}_{0,z}$ | $(100, 70, 50) E_{\text{rec,lattice}}$ |
| Lattice trap frequency | $\nu_x, \nu_y, \nu_z$ | $(65, 55, 45)$ kHz |

**Table S1. Typical operating parameters.**

| Requirement | Inequality |
|---|---|
| Ground band loading | $E_{\text{F,XODT}}, k_B T_{\text{lattice}} \ll E_{\text{rec}}$ |
| Double occupancy suppression | $J \ll U, k_B T_{\text{lattice}} \ll E_{\text{F,lattice}} \ll U$ |
| Resolved contact interactions | $\Delta U \gg 1/\tau$ |
| Lamb-Dicke | $h\nu \gg E_{\text{rec,clock}}$ |
| Doppler suppression | $8 J_x/h \ll 1/\tau$ |

**Table S2. Requirements for clock operation in a Mott-insulating regime with one atom per site.**

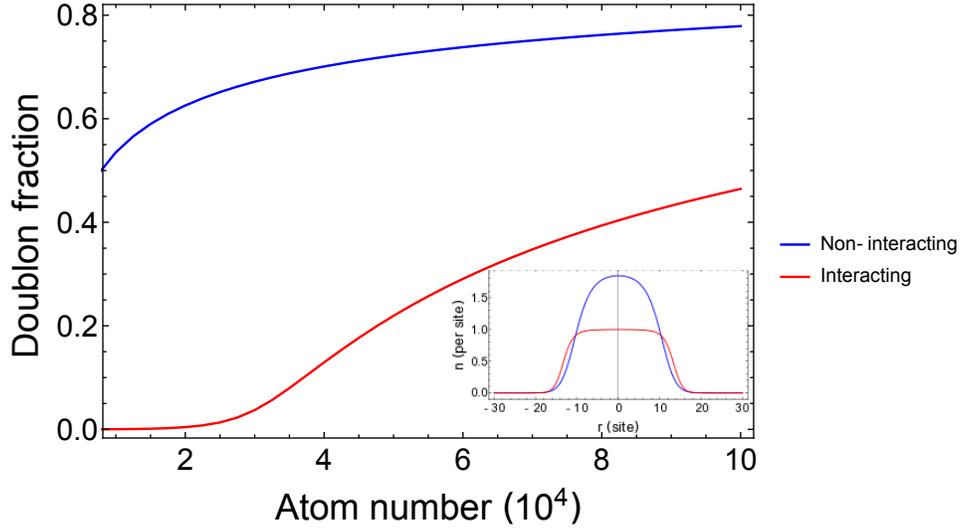

**Fig. S1.** Suppression of doubly occupied sites (doublons) in the Mott-insulating regime. A comparison of calculated doublon fraction between interacting and non-interacting cases is plotted for $T = 15$ nK and $T/T_F = 0.1 - 0.2$ in the XODT. The inset shows the calculated density profile in the lattice for $N = 1 \times 10^4$ and $T/T_F = 0.2$. For our coldest samples at $T = 15$ nK and $N = 1 \times 10^4$, used for Fig. 3 and 4, the number of doubly occupied sites is greatly suppressed as compared to calculations for the non-interacting case.

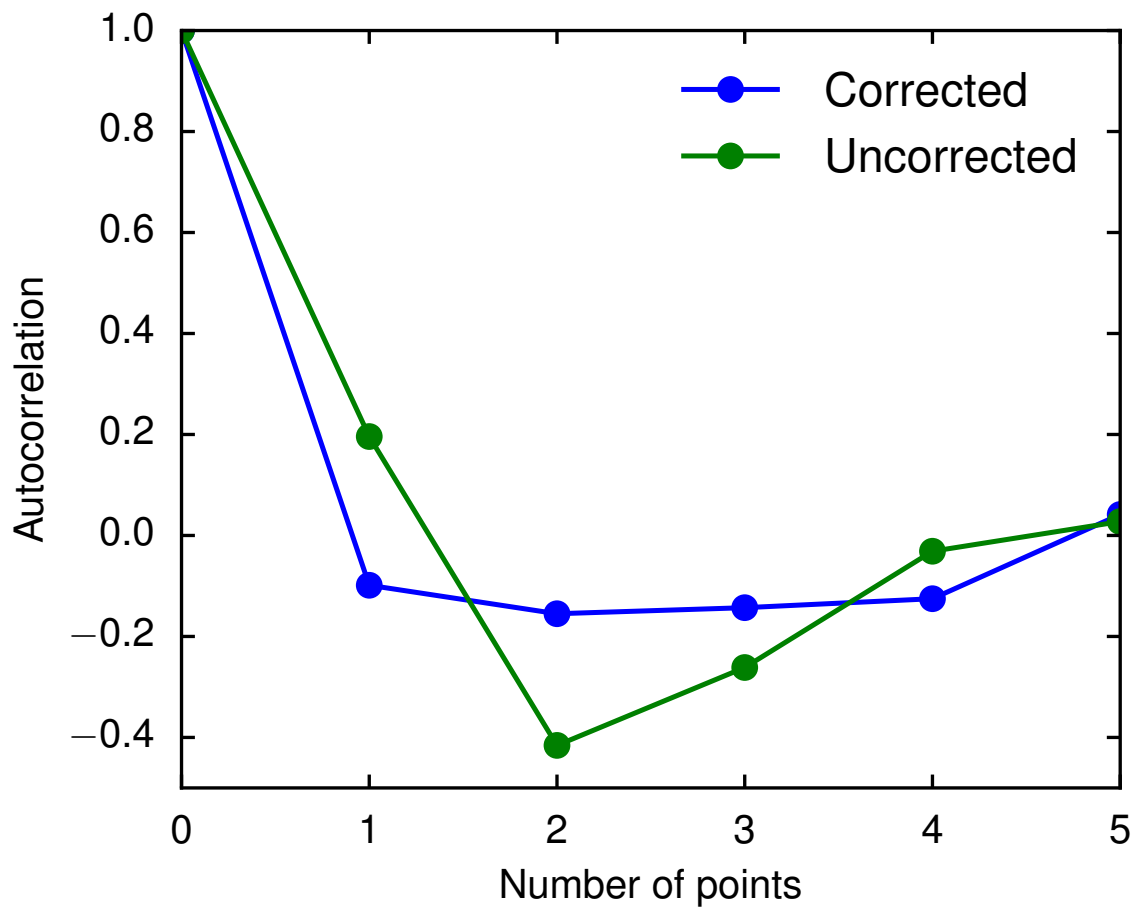

**Fig. S2. Autocorrelation plot.** Autocorrelation functions of the measured ac Stark shifts for clock laser servo frequencies uncorrected (green) and corrected by the measured error signal (blue).

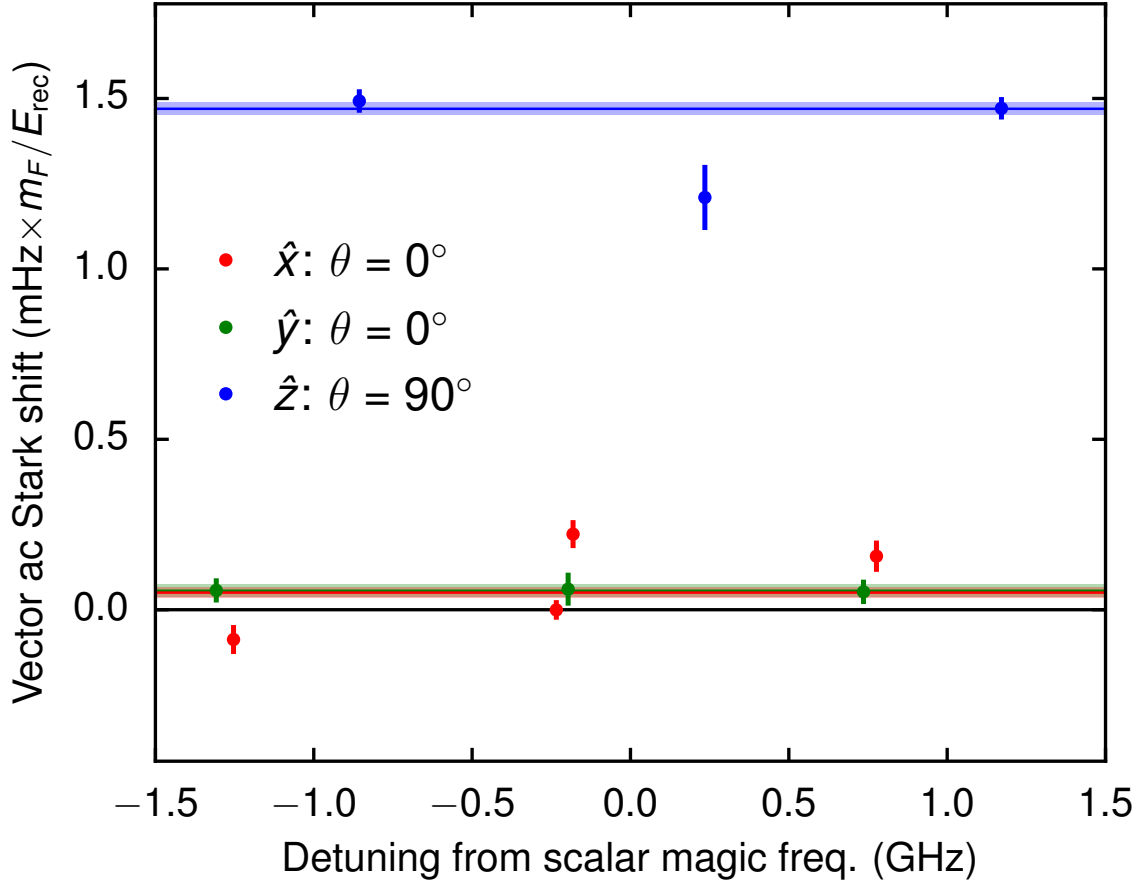

**Fig. S3. Vector ac Stark shift.** Measuring the vector ac Stark shift for the horizontal ($\hat{x}, \hat{y}$) and vertical ($\hat{z}$) configurations. The vector shift from the vertical lattice beam is sensitive to a small ellipticity in the polarization because $\hat{e}_k \cdot \hat{e}_B = 1$, while the horizontal beams with $\hat{e}_k \cdot \hat{e}_B = 0$ benefit from additional suppression of the vector shift.

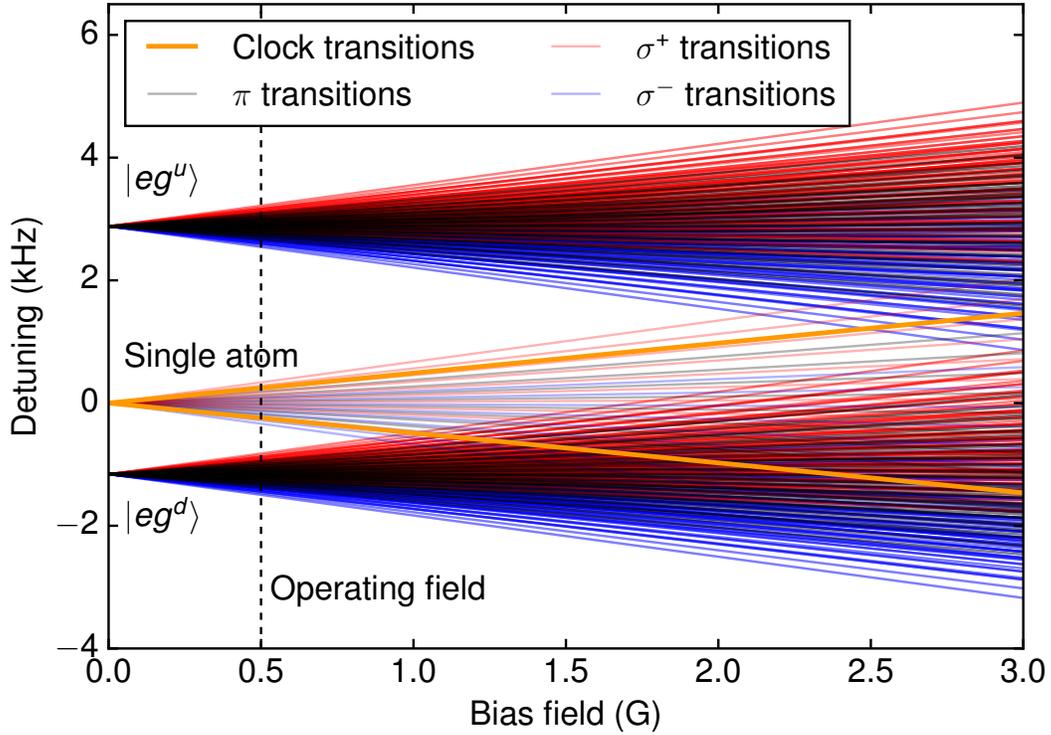

**Fig. S4. All transitions from singly and doubly occupied sites.** Detunings (relative to the unperturbed clock transition) for all transitions that can be driven on singly and doubly occupied lattice sites, shown for typical trap depths. Fig. 2B in the main text is a simplified version of this figure, with the shaded regions covering the range of all lines, and the solid lines showing the transition frequencies for only the states with $m_F = \pm 9/2$.

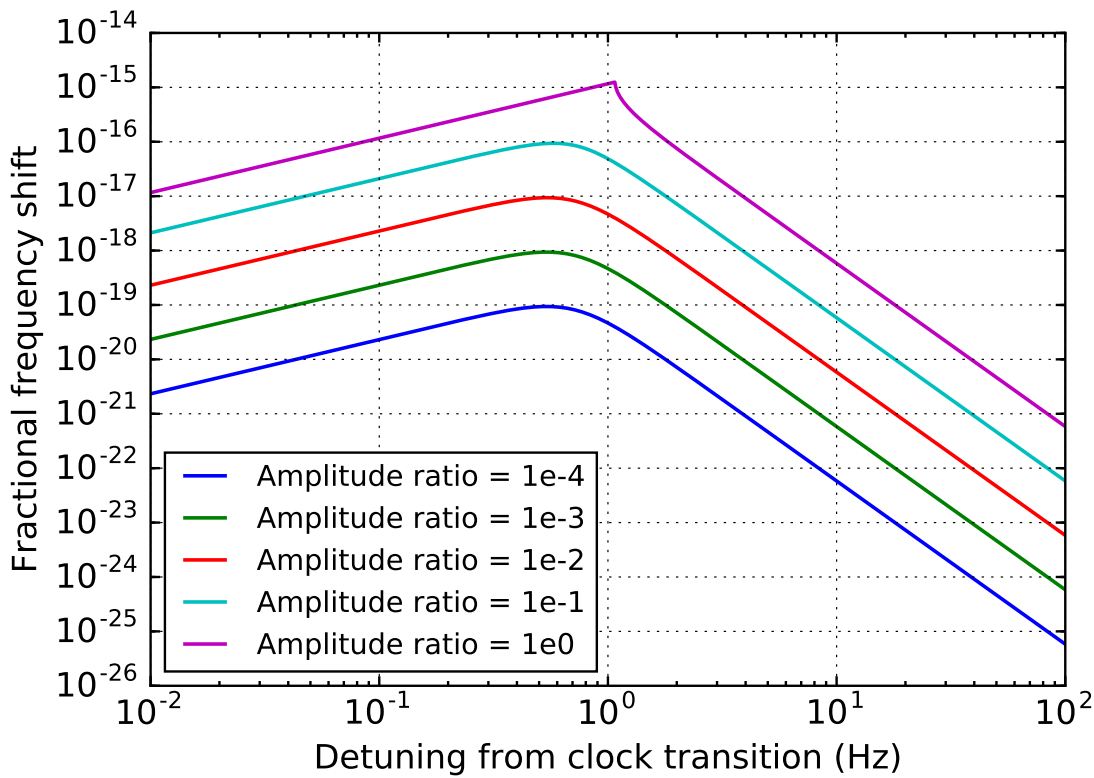

**Fig. S5. Line pulling.** Systematic shifts due to line pulling from a residual line. Though the linewidth of the residual line will vary depending on what transition is being driven, we perform this calculation for 1 Hz linewidths to give a reasonable upper bound. The amplitude ratio indicates the amplitude of the residual line relative to that of the clock transition.

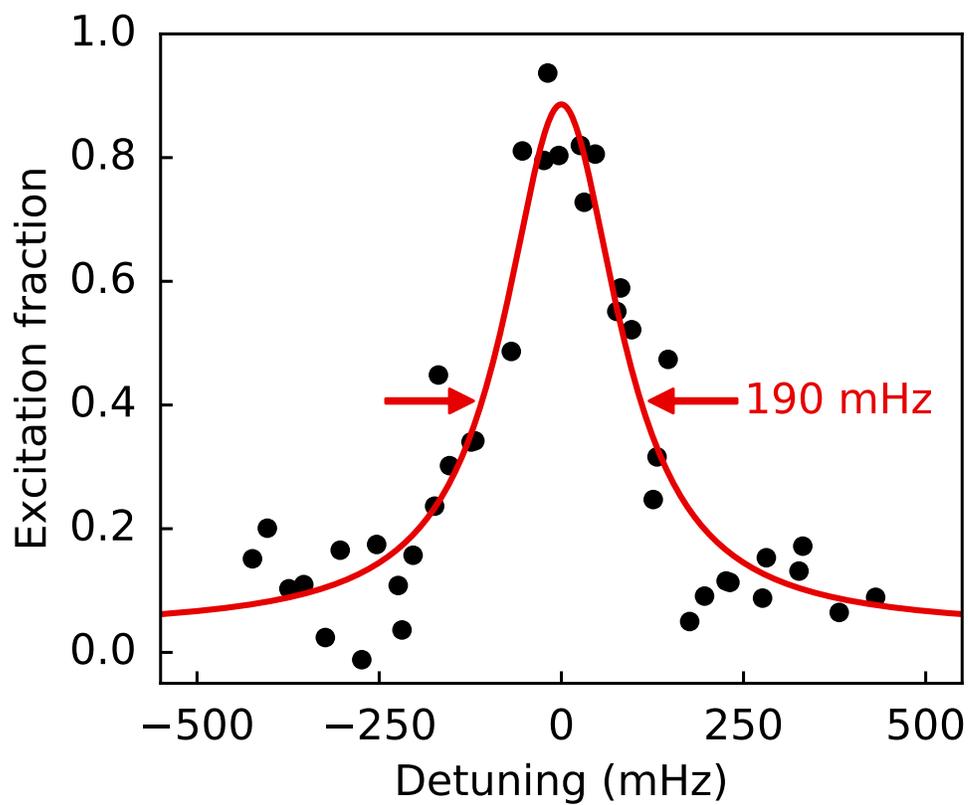

**Fig. S6. Narrow-line Rabi spectroscopy.** Rabi spectroscopy data for a 4 s $\pi$-pulse time, showing a 190(20) mHz Fourier-limited linewidth, taken with $m_F = 9/2$ and rescaled by the relative spin population.